\documentclass[aip,reprint]{revtex4-1}

\draft 

\usepackage[utf8]{inputenc}
\usepackage[T1]{fontenc}
\usepackage{mathptmx}
\usepackage{amsmath}
\usepackage{amsfonts}%
\usepackage{amssymb}
\usepackage{graphicx} 
\usepackage{blindtext}
\usepackage{nicefrac}

\begin{document}


\title{Structural and dynamical properties of gel networks} 



\author{M. Gimperlein}
\email[]{matthias.gimperlein@fau.de}
\author{M. Schmiedeberg}
\email[]{michael.schmiedeberg@fau.de}
\affiliation{Institute of theoretical physics 1, FAU Erlangen-Nuremberg}

\date{\today}

\begin{abstract}
The competition of depletion attractions and longer-ranged repulsions
between colloidal particles in colloid-polymer mixtures leads to the
formation of heterogeneous gel-like structures. For instance, gel
networks, i.e., states where the colloids arrange in thin strands that
span the whole system occur at low packing fractions for attractions
that are stronger than those at the binodal line of equilibrium
liquid-fluid phase separation. By using Brownian dynamics simulations we
explore the
formation, structure, and ageing dynamics of gel networks. We determine
reduced network that focus on the essential connections in a gel
network. We compare the observed properties to those of bulky gels or
cluster fluids. Our results demonstrate that both the structure as well
as the (often slow) dynamics of the stable or meta-stable heterogenous
states in colloid-polymer mixtures possess distinct features on various
length and time scales and thus are richly divers.
\end{abstract}

\pacs{82.70.Dd,82.70.Gg,83.80.Kn}

\maketitle 

\section{Introduction}
\label{sec1}
In colloid-polymer mixtures the effective interaction between the
colloids can be modelled by repulsions at very small and longer
distances due to screend Coulomb interactions \cite{dlvo1,dlvo2}. At
intermediate distances where there only is a small gap between the
colloids that is too small for the polymers, depletion effects lead to
an effective attraction between the colloids that is well described by
the AO-approach \cite{ao,vrij,binder}. The competition between the
repulsive and the attractive interactions can lead to formation of
complex ordered structures \cite{edelmann,oguz}. However, here we are
interested in gels that occur in such a system
\cite{verhaegh,lu,speck1,speck2,cho,archer,toledano,zhang,mani,helgeson,kohl,doorn,kohl2}.

In systems whith a strong attraction but no or only a small repulsion at
a longer range clumpy gels are observed that occur when the system
demixes into clusters at a high density and dilute pores in between
\cite{verhaegh,lu,speck1,speck2,cho}. In contrast, in case of the full
competition between attraction and repulsion thin strands are observed
that form network structures
\cite{archer,toledano,zhang,mani,helgeson,kohl,doorn,kohl2}. While in
clumpy gels the slowdown of dynamics seems to occur due to an arrested
fluid-liquid phase separation \cite{lu,speck1,speck2}, in gel networks
the slowdown usually
is attributed to a percolation transition
\cite{toledano,mani,helgeson,kohl,kohl2}. More specifically, the
slowdown and the onset of ageing along with resulting phenomena like
syneresis have been related to a directed percolation transition in
space, i.e., the strands not only have to span the whole system but they
have to do so in a directed way without any loops \cite{kohl,kohl2}.

Note that in colloid-polymer mixtures both, clumpy gels as well as gel
networks can be observed \cite{archer,zhang,mani,helgeson}. In
equilibrium at sufficiently large packing fraction clumpy gels are
observed in a similar way as for sticky particles without longer ranged
repulsions, i.e., gelation in the sense of dynamically arrested states
starts when the spinodal line of the fluid-liquid phase separation is
crossed towards larger attractions, i.e., clumpy gels only occur for
attractions larger than those at the binodal line \cite{zhang,speck2}.
\begin{figure}
\includegraphics[width=\columnwidth]{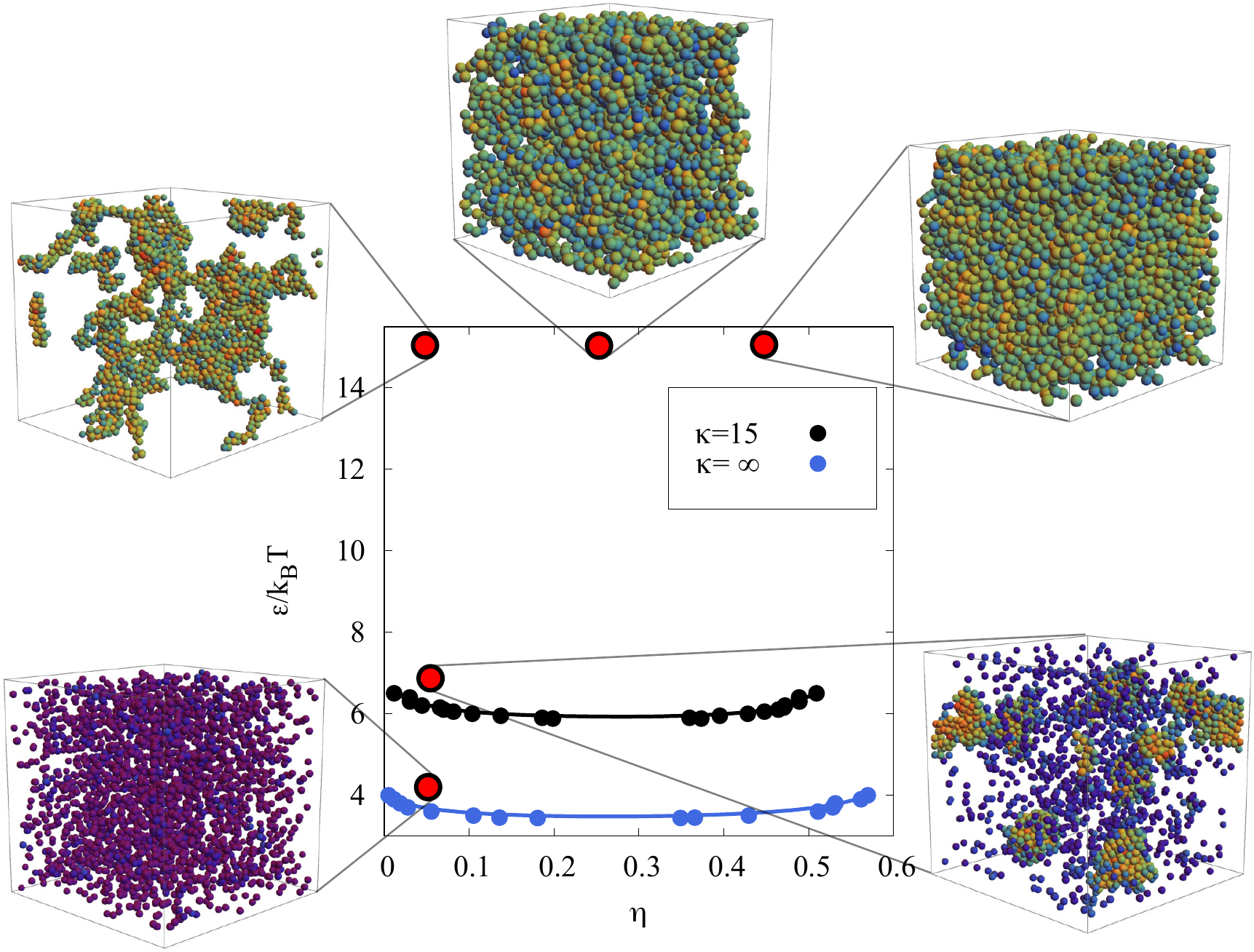}
\caption{\label{fig1-1} Phase diagram of the considered model system
that resembles a colloid-polymer mixture. Details of the model are given
in Sec \ref{sec2.1}. The black line shows the binodal line of the
gas-liquid phase separation for $\kappa$=15. For comparison, the blue
line corresponds to the binodal line for purely attractive particles
(i.e., for $\kappa=\infty$). The snapshots shown as insets illusutrate
typical configurations at the given parameter sets (all for
$\kappa$=15). Note that our system is similar to the one studied in
\cite{speck2} and our binodal lines are in agreement with the one found
in \cite{speck2}. However, no heterogeneous states have been reported in
\cite{speck2} above the binodal line.}
\end{figure}
However, in dilute systems below a crossover packing fraction
\cite{helgeson}
with sufficient longer-ranged repulsion the colloids organize in a filigree
of thin strands \cite{toledano,mani,helgeson,kohl,kohl2}. In this
article we refer to these structures when we use the term gel network.
Obviously stable or at least meta-stable gel networks only occur for
attractions that are stronger than at the equilibrium binodal line as
otherwise the strands are not stable. As already mentioned, the dynamics of
a gel network slows down and ageing sets in at the directed percolation
transition \cite{toledano,mani,helgeson,kohl,kohl2}. In summary, for
packing fractions above a crossover packing fraction the
slowdown of dynamics is related to a liquid-fluid phase separation and
results in clumpy gels. Percolation transitions are not important for
these systems as they are usually all percolated \cite{speck2}. In
contrast, for densities below the crossover packing fraction the slowdown leading to gel networks is related to a percolation transition that then usually occurs for attractions above the binodal line (see, e.g., \cite{helgeson} for a complete picture).

We employ Brownian dynamics simulations to study the structure and
dynamics in different states of a gel-forming colloid-polymer mixture.
Note that the multiscale dynamics of a clumpy gel has previously been
studied in \cite{cho} but up to our knowledge we present an extensive
analyzes of the dynamics on different lengths scales in gel networks for
the first time.

As an overview of the different states considered in this article, we
present an equilibrium phase diagram in Fig. \ref{fig1-1} for orientation.
We employ the model that previously has already been considered in
\cite{speck2}. Details of the model are given in Sec. \ref{sec2}.
We are interested in the situation with competing short-range
attractions and longer-ranged repulsions. The corresponding binodal line
of demixing is indicated by a black line in Fig. \ref{fig1-1}, e.g., for
smaller attractions the system always is fluid while for stronger
attractions different stable or meta-stable cluster fluids and gel-like
states are observed (see insets in Fig. \ref{fig1-1}). For example, close
to the binodal line at small densities clusters are observed that can
still move around as in a fluid as long as the system is not percolated.
For stronger attraction percolated gel network structures occur at small
packing fractions while for larger densities clumpy gels are observed.
Our main interest in this work are the properties of the gel networks
and how they differ from the cluster fluids or bulky gels.

Note that while we consider the model employed in \cite{speck2} our
findings are in conflict to the claims in \cite{speck2}. In
\cite{speck2} only systems with weak attraction, i.e., below or at the
binodal line are considered. Nevertheless it was claimed that no other
behavior could occur above the binodal line and especially that there
was no transition between a cluster fluid and a gel network above the
binodal line \cite{speck2}. Our observance of a cluster fluid above the
binodal already already proves that in \cite{speck2} the findings from
below the binodal lines have wrongly been generalized to systems above
the binodal line. Our main focus lies on structures with stronger
attractions, i.e., above the binodal line.

The article is organized as follows: In Sec. \ref{sec2} we first
introduce the model system, before we describe the Brownian dynamics
simulations, and our methods of analyzes. In Sec. \ref{sec3} we
present our simulation results concerning the thermalization process,
the structure, as well as the dynamical properties of gel networks.
Finally, we conclude in Sec. \ref{sec4}.
\section{Model and simulation details}
\label{sec2}
\subsection{Model details}
\label{sec2.1}
\begin{figure}
\includegraphics[width=\columnwidth]{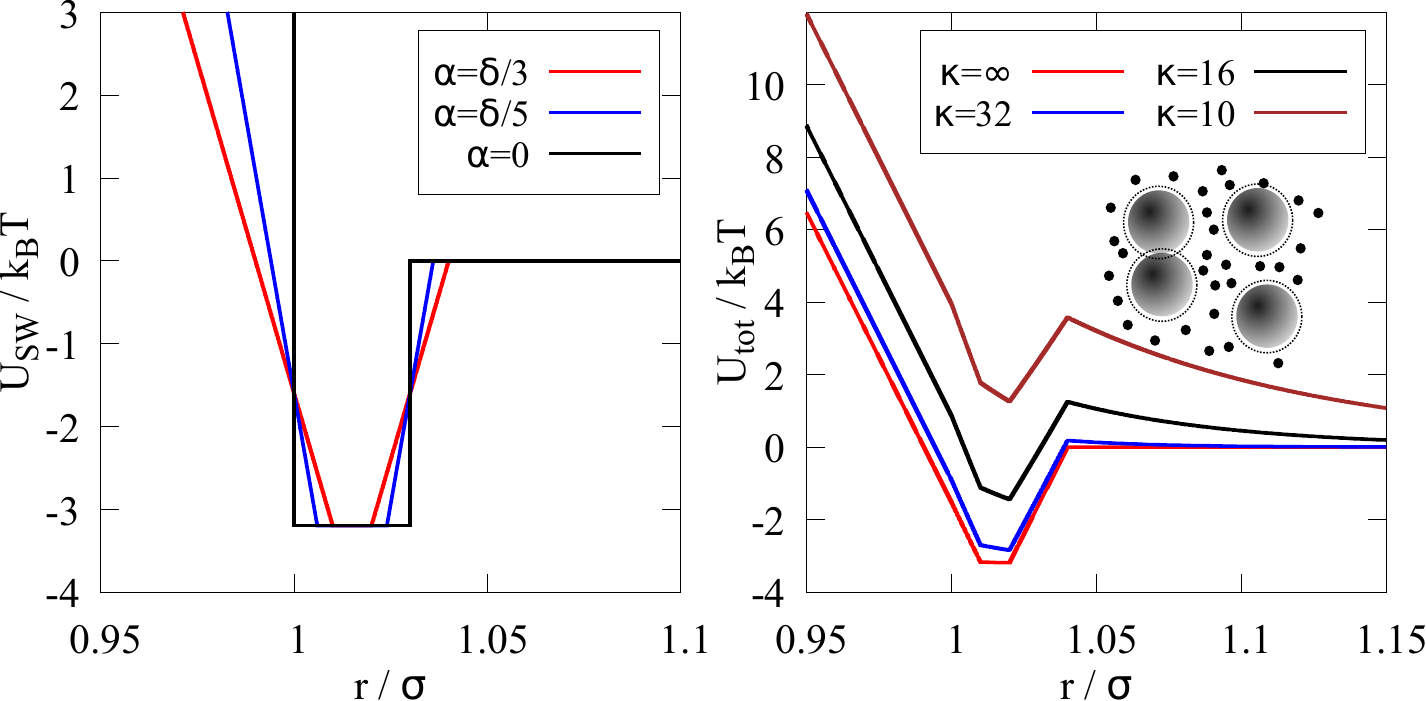}
\caption{\label{fig1.2-1} The overall potential (shown in the right
panel for different screening lengths $\kappa^{-1}$) is a sum of a repulsive
Yukawa potential $U_{YK}$ and a modified square-Well-Potential $U_{SW}$
(depicted on the left hand side) with width $\delta=0.03\sigma$. To
avoid infinite forces the jumps in the square well potential are
smoothened by introducing a parameter $\alpha$ as given in the text. In
the rest of the article $\alpha=\frac{\delta}{5}$ is used. In the
examples shown here, the strength of the attraction always is $\epsilon=3$.
}
\end{figure}

We perform Brownian dynamics simulations in a system closely related to
the one presented in \cite{speck2} consisting of $N$ collidal particles
whose radii are drawn from a Gaussian distribution, such that the
polydispersity of the system is 5\%. The mean diameter of the particles
is called $\sigma$ and the sum of radii of particles $i$ and $j$ is
$\sigma_{ij}$. The short range attraction is modeled via a modified
square-well potential $U_{\text{SW}, ij}(r)$ of width
$\delta=0.03\sigma$ and depth $\epsilon$, the longer ranged repulsion is
given by a Yukawa potential $U_{\text{YK}, ij}(r)$ which models screened
electrostatic repulsion. The overall interaction potential shown in Fig.
\ref{fig1.2-1} is the sum
$U_{\text{tot}, ij}(r)=U_{\text{YK}, ij}(r)+U_{\text{SW}, ij}(r)$ where
\scalebox{.93}{\parbox{\columnwidth}{
\begin{align*}
U_{\text{YK}, ij}(r)&=C \left( \frac{2}{2+\kappa
\sigma_{ij}}\right)^2\left(\frac{\sigma_{ij}}{r}\right)\text{exp}[-\kappa(r-\sigma_{ij})],\\
U_{\text{SW},ij}(r)&=\begin{cases}
-\frac{\epsilon}{2\alpha}r+\frac{\epsilon(1-\alpha)}{2\alpha} & \text{if
} r<\sigma_{ij}+\alpha\\
-\epsilon & \text{if } \sigma_{ij}+\alpha \leq r \leq
\sigma_{ij}+\delta-\alpha\\
\frac{\epsilon}{2\alpha}r - \frac{\epsilon(1+\delta+\alpha)}{2\alpha}
&\text{if } \sigma_{ij}+\delta-\alpha < r < \sigma_{ij}+\delta+\alpha \\
0 & \text{else} .
\end{cases}
\end{align*}}}
The additional parameter $\alpha$ flattens the wells of the square-well
potential, which is necessary to avoid inifite forces when performing
Brownian dynamics simulations. In our simulations we choose
$\alpha=\frac{\delta}{5}$. The screening length $\kappa^{-1}$, which
represents the strength of the repulsive force, can be tuned by
modifying salt concentration in experiments \cite{kohl}. $\kappa=\infty$
represents the case of a purely attractive potential without repulsive
forces. The parameter $C$ is chosen as 200$k_BT$. The cut-off distance
is chosen to be $\frac{r_{\text{Cut}}}{\sigma}=1+\frac{4}{\kappa}$ as in
\cite{speck2}. In summary the whole system can be characterized by choosing a triplet of parameters $(\epsilon, \kappa, \eta)$, where
$\eta=\frac{\pi}{6}\sigma^3\frac{N}{L^3}$ is the packing fraction of the
system with the box size $L$. Note that in the following $\epsilon$ is
used in units of $k_BT$ and $\kappa$ in units of $\sigma^{-1}$.
\subsection{Brownian Dynamics}
\label{sec2.2}
We employ Brownian dynamics simulations where the motion of colloidal
particles in a solvent is determined by numerically integrating the overdamped
Langevin-equation for particle $j$
\begin{align*}
\gamma \frac{d}{dt}\vec{r}_j=\vec{F}_{\text{int}}+\vec{F}_{\text{th}},
\end{align*}
where $\gamma$ is the friction constant, $\vec{F}_{\text{int}}$ models
the effective force between colloidal particles as given by the pair
interaction potential introduced in the pevious subsection, and
$\vec{F}_{\text{th}}$ denotes random forces due to thermal fluctations.
It is $\left\langle \vec{F}_{\text{th}}\right\rangle=\vec{0}$ and
$\left\langle F_{\text{th},i}(t)F_{\text{th},j}(t')\right\rangle=2\gamma
k_\textnormal{B} T\delta_{ij}\delta(t-t')$. The time steps in our
simulations are $\Delta t=10^{-5} \tau_B$ with the Brownian time
$\tau_B=\frac{\sigma^2\gamma}{4k_\textnormal{B} T}$. The calculation of
forces was done using a combination of the Verlet-list algorithm and the
linked-cell algorithm to reduce computation time\cite{allen}.  For $\eta=0.05$ boxes of size $30\sigma\times 30\sigma\times 30\sigma$ with
periodic boundary conditions are used and filled with colloids until the
considered packing fraction is reached, i.e., we simulate with $N=2578$
particles. For $\eta=0.25$ we use boxes of size $20\sigma\times 20\sigma\times 20\sigma$ and $N=3819$ particles.
\subsection{Reduced Networks}
\label{sec2.3}
\begin{figure}
\includegraphics[width=\columnwidth]{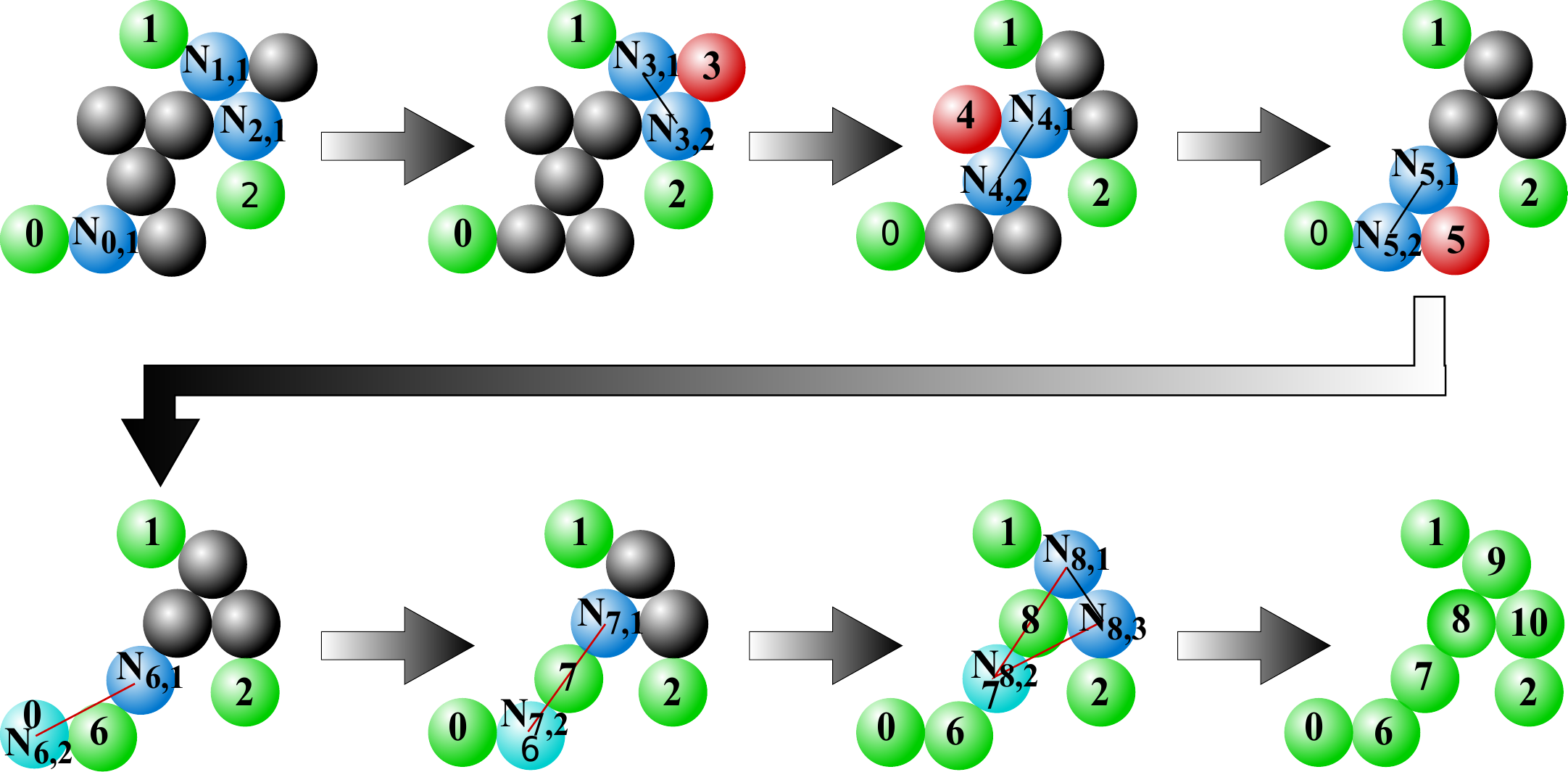}
\caption{\label{fig2.3-1} Schematical representation of the algorithm
developped to determine a reduced network. Starting with the particle
$i$ which has the minimal coordination number. If the neighbors
$N_{i,j}$ are still connected when the chosen particle $i$ is removed it
will be removed, otherwise it is kept and marked as visited. Removed
particles are shown in red, visited and kept particles in green. The
algorithm continues until every particle has been visited.}
\end{figure}
Complex network structures occur in some regions of the phase diagram
(see e.g. Fig. \ref{fig1-1}). To extract the essential part of these
networks, we determine so-called \textit{reduced networks}. The key idea
is to remove as many particles as possible without destroying
connections within the network. The algorithm is illustrated in Fig.
\ref{fig2.3-1} and works as follows
\begin{itemize}
\item[\textbf{0.}] Remove all single particles which have no neighbors.
\item[\textbf{1.}] Choose the particle $i$ with the minimal number of
neighbors.
\item[\textbf{2.}] Check whether particle $i$ was already chosen before.
\begin{itemize}
\item[\textbf{Yes:}] Keep particle $i$ and start again at step 1
(ignoring particle $i$ from now on).
\item[\textbf{No:}] Calculate all neighboring particles $N_{i,j}$ for
the chosen particle $i$ and check whether these are still connected over
a certain amount of steps (we choose 3 steps) if $i$ would be removed.
\begin{itemize}
\item[\textbf{Yes:}] Delete particle $i$.
\item[\textbf{No:}] Keep particle $i$ and mark it as visited.
\end{itemize}
\end{itemize}
\item[\textbf{3.}] Start again at step \textbf{1}, until every particle
has been marked as visited.
\item[\textbf{4.}] When every particle was visited start the algorithm
again with the reduced network as initial configuration until no more
particles are removed during the whole procedure.
\end{itemize}
The results of this analysis is shown later within this work, e.g., Fig.
\ref{fig3.2-2}.

\section{Results}
\label{sec3}
\subsection{Initial evolution towards heterogeneous structure}
\label{sec3.1}
\begin{figure}
\includegraphics[width=\columnwidth]{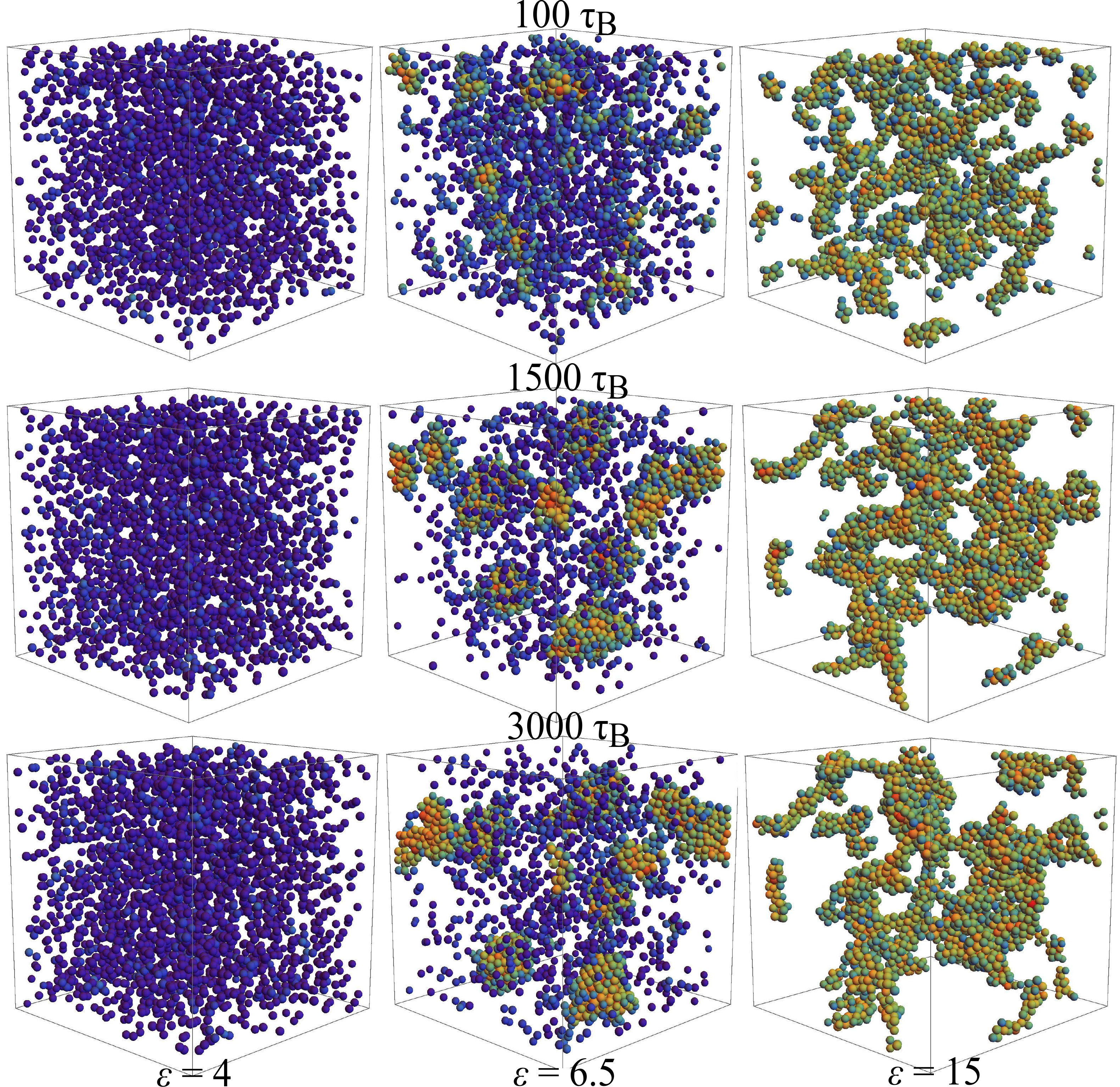}
\caption{\label{fig3.1-1} Evolution of the system for an inverse screening length $\kappa=15$ at a packing fraction $\eta=0.05$ for attraction strength
$\epsilon=4$ (left column), $\epsilon=6.5$ (center column), and
$\epsilon=15$ (right column). The colors indicate the number of nearest
neighbors each particle has, ranging from blue for no neighbors to red
for 12 neighbors.}
\end{figure}
We first study how the particles that are initially placed at random
positions organize to form the heterogeneous structures that we are
interested in. In Fig. \ref{fig3.1-1} we show snapshots of the evolution
for different systems at a low packing fraction of $\eta=0.05$ and
different strength of attraction $\epsilon$ after $100\tau_B$,
$1500\tau_B$, and $3000\tau_B$ where $\tau_B$ is the Brownian time
introduced in Sec. \ref{sec2.2}. The case shown in the left column of
Fig. \ref{fig3.1-1} with an attraction strength $\epsilon=4$ is on the
fluid side of the binodal line. Therefore, no complex structures occur.
On the opposite side of the binodal line, i.e., where phase separation
is expected in equilibrium, heterogeneous structures are observed.
Depending on the attraction strength different regimes can be observed.
To be specific, close to the binodal line, e.g., for $\epsilon=6.5$,
unconnected clusters of particles are found (center column). As these
clusters can move freely, we term this state a cluster fluid. Far above
the binodal line, e.g., for $\epsilon=15$ as shown in the right column,
a gel network with percolating strands is formed. Furthermore, unbound
particles are rare in such gel networks.

The initial structure formation occurs within a short time. Already
after $100\tau_B$ (top row of Fig. \ref{fig3.1-1}) the most important
features of a structure are visible. Not that between $100\tau_B$ and
$1500\tau_B$ (center row) there are still structural changes visible in
case of the heterogeneous structures. However, between $1500\tau_B$ and
$3000\tau_B$ no significant changes of the type of structure is visible
from the snapshots, i.e., the system seems to relax towards a
meta-stable structure. We now want to study this initial relaxation
process in more detail and characterize the evolution of the structures
quantitatively.

\begin{figure}
\includegraphics[width=\columnwidth]{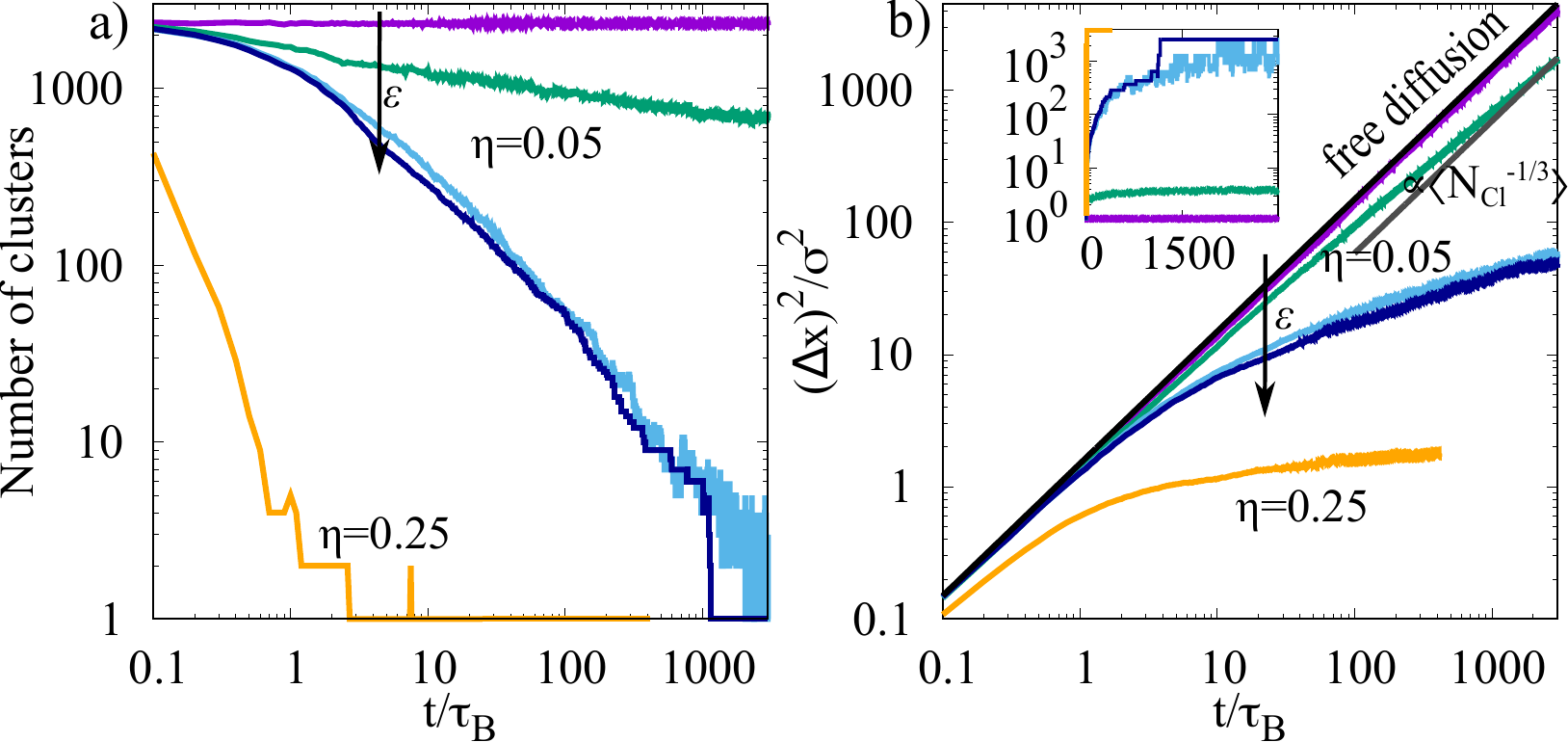}
\caption{\label{fig3.1-2} (a) Number of non-connected clusters and (b)
mean squared displacement $(\Delta x)^2/\sigma^2$ in the system during
the inital relaxation. The inset in (b) shows the mean cluster size. The
light-orange curve is a clumpy gel for $\eta=0.25$ and $\epsilon=15$
while all other curves denote systems at a low packing fraction
$\eta=0.05$ and various attraction strengths ranging from a fluid with
$\epsilon=4$ (purple), over a cluster fluid with $\epsilon=6.5$ (green),
to percolating gel networks with $\epsilon=9$ and $\epsilon=15$
(light-blue and dark-blue, respectively). Note that the colors are the
same as specified in the legend of Fig. \ref{fig3.2-1} and are used
throughout the rest of the article for all low density systems. The
black line in (b) denotes free diffusion with $(\Delta
x)^2/\sigma^2=3t/(2\tau_B)$ and the grey curve indicates $(\Delta x)^2/\sigma^2=3t/(2\tau_B)\left\langle N_{\textnormal{Cl}}^{-1/3}\right\rangle$, where $\left\langle
\cdot \right\rangle$ indicates the average over all particles and $N_{\textnormal{Cl}}$ is the size of the cluster around a particle at
$3000\tau_B$ for the cluster fluid ($\epsilon=6.5$).}
\end{figure}

Fig. \ref{fig3.1-2}{a) shows the number of clusters forming in a system
and in the inset of Fig. \ref{fig3.1-2}{b) the mean cluster size is
plotted. For low packing fraction $\eta=0.05$ the evolution strongly
depends on the attraction strength. Below the binodal line, i.e., in the
fluid phase shown by purple lines, particles hardly attach to each other
such that the number of clusters remains large and the mean cluster size
small. For the cluster fluid in the phase separated part of the phase
diagram close to the binodal line, represented with green curves, the growth of clusters can be observed that significantly
slows down for larger times though some slow evolution is still ongoing
at the end of this initial simulation. The gel networks depicted with
light-blue and dark-blue curves slowly relax towards percolated states,
i.e, finally all particles are part of one connected (network)
structure. For comparison, the case of a clumpy gel with a packing
fraction of $\eta=0.25$ and a strong attraction of $\epsilon=15$
(light-orange curves) is shown. It percolates much faster, namely within
a few Brownian times.

In Fig. \ref{fig3.1-2}(b) the mean-squared displacement during the
initial relaxation process is shown. In a fluid phase diffusive behavior
is expected in the long time limit. Indeed, for the fluid state below
the binodal line, the particles move as expected from free diffusion
which we indicate by a black line that lies on top of the purple curve.
In case of a cluster fluid the mean-squared displacement can be roughly
estimated by considering diffusion of clusters whose size corresponds to
the mean cluster size that we observe at $3000\tau_B$. If the clusters
had a spherical shape one would expect that the diffusion constant is
lowered by a factor $\left\langle N_{\textnormal{Cl}}^{-1/3}\right\rangle$ (see grey line in Fig. \ref{fig3.1-2}(b)). We indeed find that the mean-squared displacement of the cluster fluid approaches this estimated diffusive behavior.

In contrast, the mean-squared displacement of the particles in gel
networks (light-blue and dark-blue curves) flattens significantly. As a
consequence, it is still possible that the mean-squared displacement of
gel networks will reach a diffusive limit at a much longer time.
However, on the timescale accessible for our simulations, the system
behaves subdiffusively. Therefore, the dynamics slows down significantly
but is not completely arrested. The remaining dynamics can be considered
as ageing process.
\subsection{Structure}
\label{sec3.2}
As seen in the previous section the evolution of gel networks slows down
dramatically after an initial thermalization period. Note that the
system is not in perfect equilibrium and not static after the
thermalization. However, changes of the structure become very slow and
will be considered as ageing dynamics of an effectively meta-stable
state. In this subsection, we want to study the structure of the
heterogeneous states after the initial relaxation dynamics.

\begin{figure}
\includegraphics[width=\columnwidth]{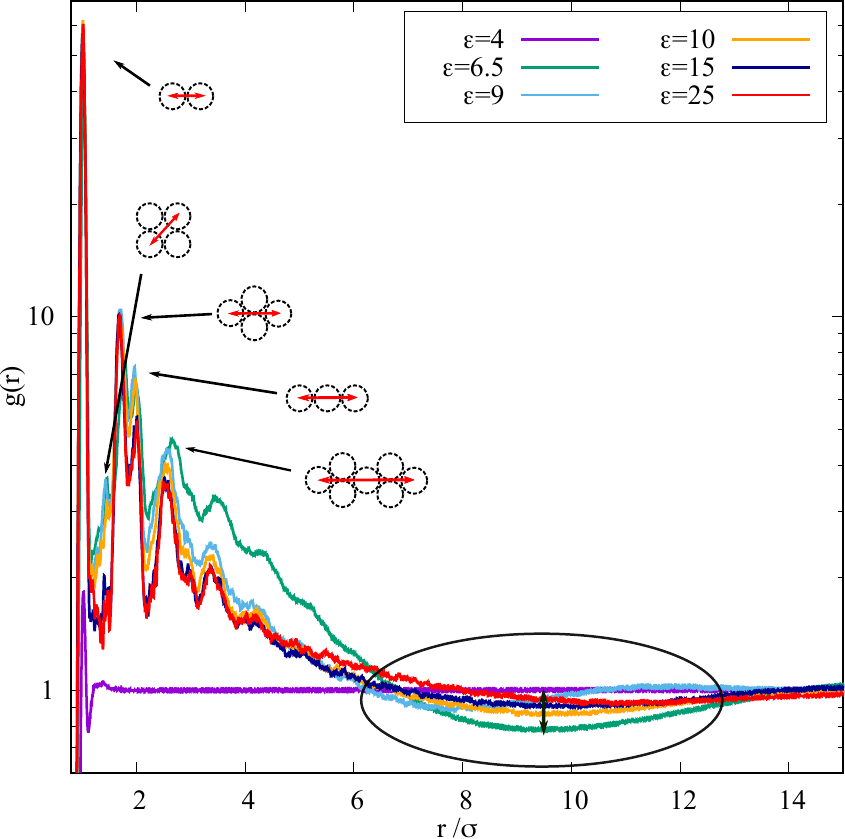}
\caption{\label{fig3.2-1} Radial pair correlation functions $g(r)$ for
$\kappa=15$, $\eta=0.05$, and different attraction $\epsilon$. The
purple curve $\epsilon=4$ is taken below the binodal and shows typical
like behavior of a homogeneous fluid, while all other pair correlation
functions taken above the binodal show signitures of cluster or network
structures. Preferred particle configuration can be identified by
studying peak positions. The mean distance of the region where $g(r)<1$
can be seen as a measure of the mesh size of a network in case of the
gel networks or as typical distance to void spaces between clusters in
case of the cluster fluid.}
\end{figure}

Fig. \ref{fig3.2-1} shows the pair correlation function $g(r)$ for low
packing fractions $\eta=0.05$, a screening length given by $\kappa=15$,
and different values of $\epsilon$. Note that the data was averaged over
500$\tau_B$ after an initial thermalization time of 2500$\tau_B$. In the
homogeneous fluid below the binodal line, i.e., for $\epsilon=4$ (purple
curve) peaks are less pronounced then in the inhomogeneous structures
above the binodal line, where the peaks can be assigned to preferred
particle configurations. Close to the binodal line ($\epsilon=6.5$,
green curve) an extended cluster can be observed up to distances of
approximately $6\sigma$ followed by a void region with $g(r)<1$. For the
gel networks with larger $\epsilon$ the initial region with $g(r)>1$
decays faster but does hardly differ for different $\epsilon$. The mean
distance of the region with $g(r)<1$ can be associated to the mesh size
of the networks. As can be seen in Fig. \ref{fig3.2-1} the mesh size
increases with increasing $\epsilon$.

\begin{figure}
\includegraphics[width=\columnwidth]{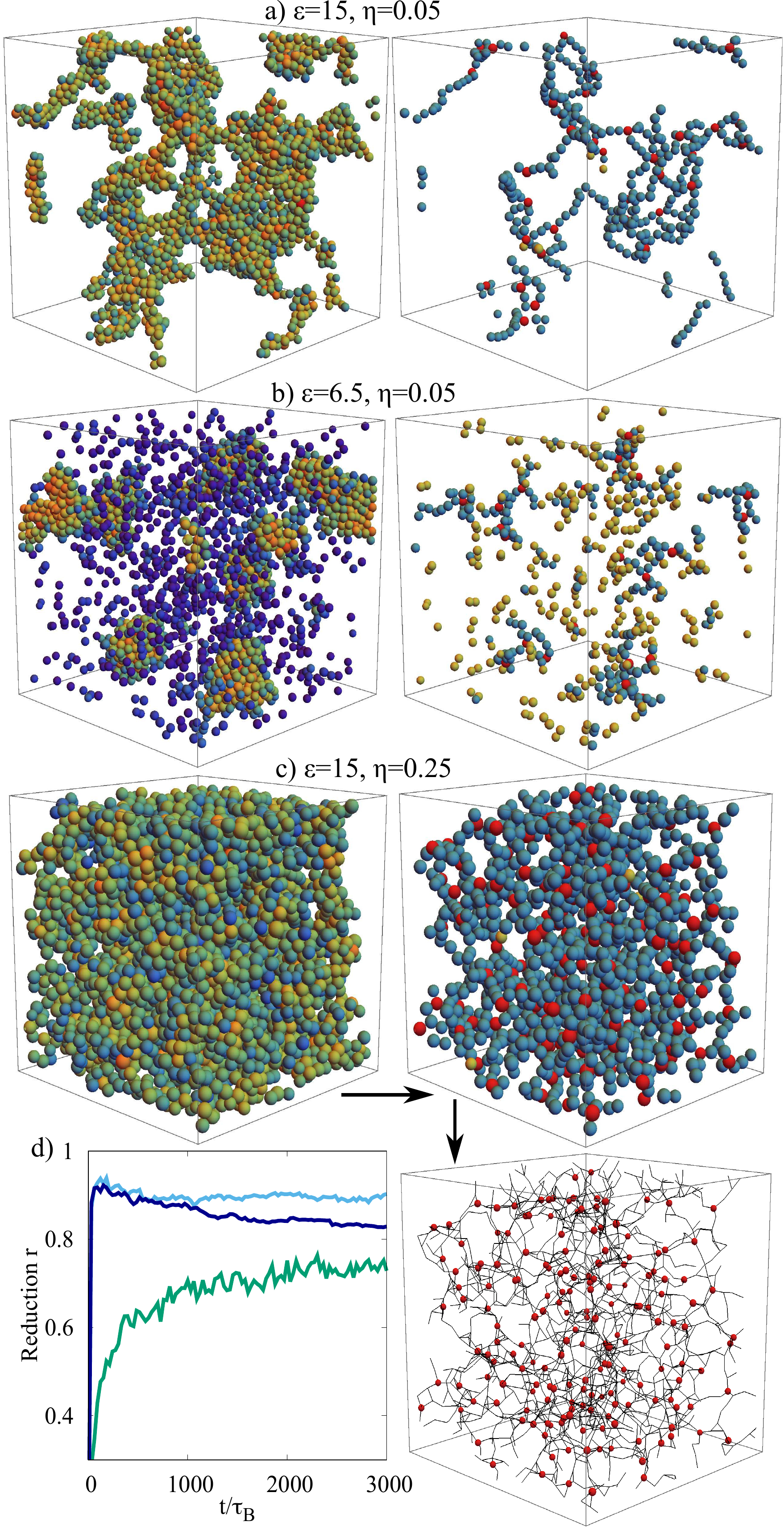}
\caption{\label{fig3.2-2} (a,b,c) Snapshots of typical heterogeneous
structures (left panels) where particles are colored by their number of
neighbors as in Fig. \ref{fig3.1-1} and the corresponding reduced networks
constructed by the method introduced in Sec. \ref{sec2.3} (right panels)
where red particles represent nodes, blue particles are colloids in
strands, and yellow particles denote the end of strings. (a) Gel
networks for $\epsilon=15$, $\kappa=15$, and $\eta=0.05$ after
3000$\tau_B$. (b) Cluster fluid for $\epsilon=6.5$, $\kappa=15$, and
$\eta=0.05$ after 3000$\tau_B$. (c) Clumpy gel for $\epsilon=15$ ,
$\kappa=15$, and $\eta=0.25$ after 400$\tau_B$ for comparison. In the
bottom panal on the right hand side we have redrawn the reduced network
structure of the clumpy gel with only showing node particles in red and
lines for connections between particles in the reduced network. (d)
Fraction of particles that are left away while constructing a reduced
network after different relaxation times. Note that unbounded particles
had already been omitted before the reduced network was constructed. The
same colors as in Fig. \ref{fig3.2-1} are used, i.e., the blue curves
are the results of gel networks and the green curve for the cluster fluid.}
\end{figure}

Next we determine typical reduced network structures as introduced in
Sec. \ref{sec2.3}. In Fig. \ref{fig3.2-2} three examples are shown,
namely a gel networks in (a), a cluster fluid in (b), and a clumpy gel
in (c). While the cluster fluid is obviously unpercolated, the shown gel
network just percolates. Note that one has to take the periodic boundary
conditions into account to see the percolation. For both, the cluster
fluid as well as the gel networks, the reduced networks only consists of
a small fraction of the original particles. In contrast, the reduced
network of a clumpy gel that we analyze in comparison in Fig.
\ref{fig3.2-2}(c) possesses a dense structure, i.e., no significant void
regions are visible.

For gel networks and the cluster fluid we show in Fig.
\ref{fig3.2-2}(d) the fraction of particles which can be removed during
the construction of a reduced network after initially all single
particles have already been removed. In gel networks for $\epsilon=9$
(light-blue) and $\epsilon=15$ (dark-blue) most particles can be
removed, i.e., most colloids are not essential for the connections in a
network. Initially, both gel networks behave similarly, but after about
1000$\tau_B$ the fraction of removed particles for $\epsilon=9$ remains
constant, while for $\epsilon=15$ it decreases. A possible
interpretation of this behavior is that for $\epsilon=15$ thinner
strands form which do not allow a high removal rate, because otherwise
connections would be destroyed.

In contrast, for cluster fluid with $\epsilon=6.5$ (green curve) fewer
particles can be removed, especially in case of smaller relaxation
times. The reason is that the cluster fluid consists of unconnected
clusters and in each of these clusters some particles have to remain in
order to denote the connections within the cluster. The increase of the
green curve in Fig. \ref{fig3.2-2}(d) results from the decrease of the
number of clusters during the initial thermalization.

\subsection{Dynamics of network structure}
\label{sec3.3}

In this subsection, we explore the dynamics of the heterogeneous
structures after an initial relaxation process. This dynamics can be
considered as ageing dynamics or as a (slow) continuation of the relaxation.

\subsubsection{Spatial correlations}
\label{sec3.3.1}
\begin{figure}
\includegraphics[width=\columnwidth]{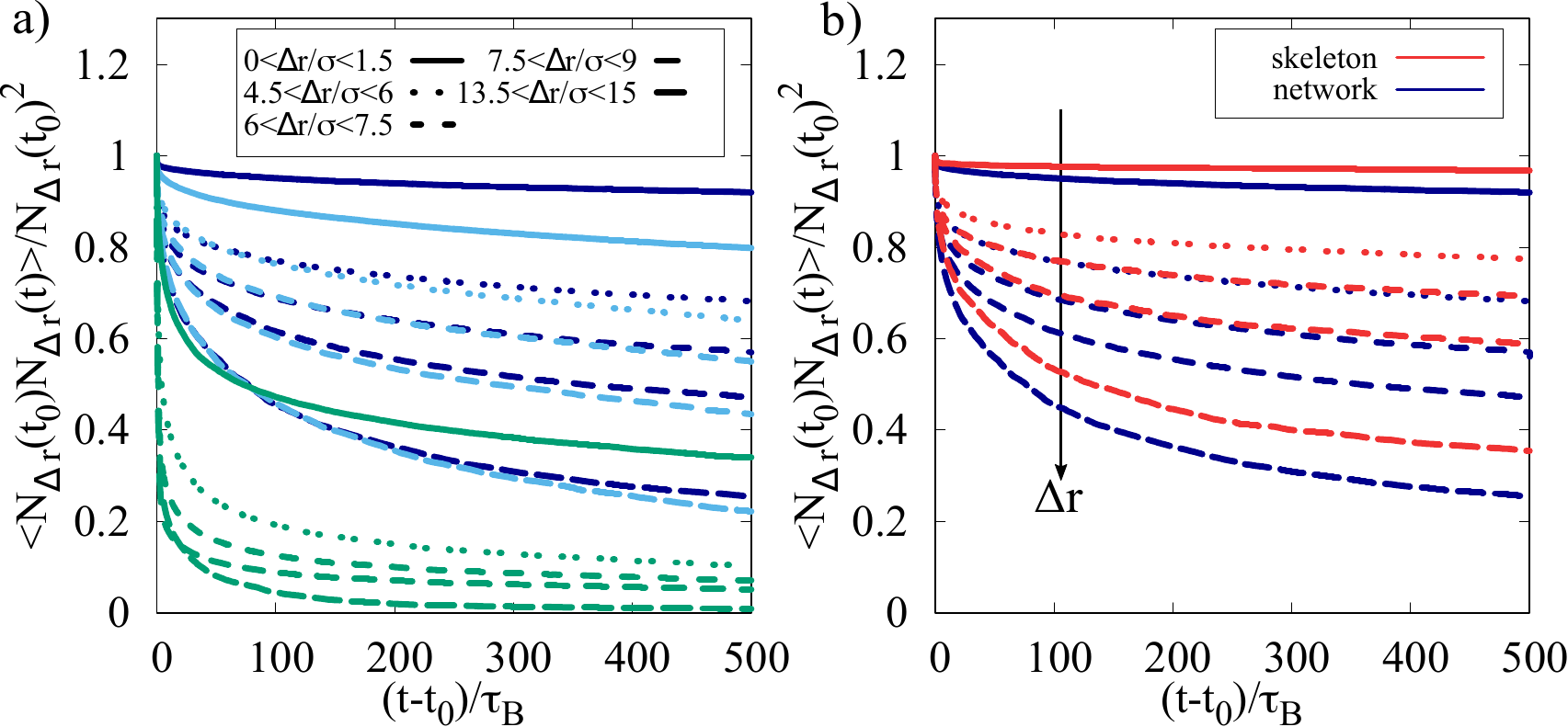}
\caption{\label{fig3.3.1-1} Correlation functions $\left\langle
N_{\Delta r}(t_0)N_{\Delta r}(t)\right\rangle$ indicating what fraction of
particles that initially are at a spatial distance $\Delta r$ from
within the intervals given in the legend stay within these intervals up
to the time $t$. A detailed definition of $N_{\Delta r}(t)$ is given in
the text. Data is averaged over at least 4 individual runs to increase
the statistics and the initial relaxation time is $t_0=3000\tau_B$. (a)
Correlation functions for gel networks (dark-blue for $\epsilon=15$ and
light-blue for $\epsilon=9$) and a cluster fluid (green,
$\epsilon=6.5$). In all cases, $\kappa=15$ and $\eta=0.05$. The line
style denotes the intervals of the spatial distance as specified in the
legend. (b) Correlation functions for the case $\epsilon=15$, function
evaluated for all particles in comparison to the correlations only
determined for the particles that are part of the reduced network
(skeleton).
The correlation of particles in the reduced network is stronger which
can be explained by boundary particles at the outside of strings which
are removed in the skeleton. These are able to leave the string and can
move easier.}
\end{figure}

For the particles that at a time $t_0$ have been at a spatial distance
$\Delta r$ from an arbitrarily chosen reference particle, we first
determine the number of particle $N_{\Delta r}(t)$ that at time $t$ have
not left a certain interval around the distance $\Delta r$. Note that
particles that during some time are closer or further away than allowed
by the chosen interval are no longer counted even if they return at a
later time. In the next step we calculate the correlation function
$\left\langle N_{\Delta r}(t_0)N_{\Delta r}(t)\right\rangle$ which by
construction is a monotonously decreasing function.

In Fig. \ref{fig3.3.1-1}(a) the correlation functions are shown for two
gel-networks (light-blue and dark-blue) as well as for the cluster fluid
(green). For the cluster fluid the correlation functions decay much
faster than for the gel networks. Only for small $\Delta r$ a
significant amount of particles stay at a similar distance which
probably are particles within the same cluster. For larger distances the
rapid decay for the cluster fluid indicates that the clusters in a
cluster fluid are unconnected and therefore can move around freely.

For the gel networks the decay is more pronounced for larger distances
indicating that some network rearrangements still are possible. The most
pronounced differences between the two gel networks occur for the case
of small distances: For particles that were neighbors in the beginning,
the correlations for $\epsilon=9$ decay faster than for $\epsilon=15$
because bonds between particles can rupture much easier in case of
smaller attractions. Interestingly, while some small differences are
visible for the different gel networks in most cases, for a distance in
the range $6\sigma$ to $7\sigma$ hardly any difference between the
correlation functions can be noticed. Note that this distance
approximately corresponds to half a mesh size as expected from Fig.
\ref{fig3.2-1}. For shorter distances particles that disconnect or
connect to strands probably make a difference, while the difference for
larger distances is due to network
rearrangements that occur on the length scale of the mesh size or even
larger length scales.

It is also possible to calculate the spatial correlation functions for
the particles which are part of the corresponding reduced network. Fig.
\ref{fig3.3.1-1}(b) shows for $\epsilon=15$ a comparison between the
correlations in the whole network (dark-blue curves) and for the
skeleton network (red curves). We notice that the correlation
function of the skeleton decays slower than the corresponding whole
network function. This is independent of the distance indicating that
particles of the reduced networks on average are more stable than
average particles of the whole structure.

\begin{figure}
\includegraphics[width=\columnwidth]{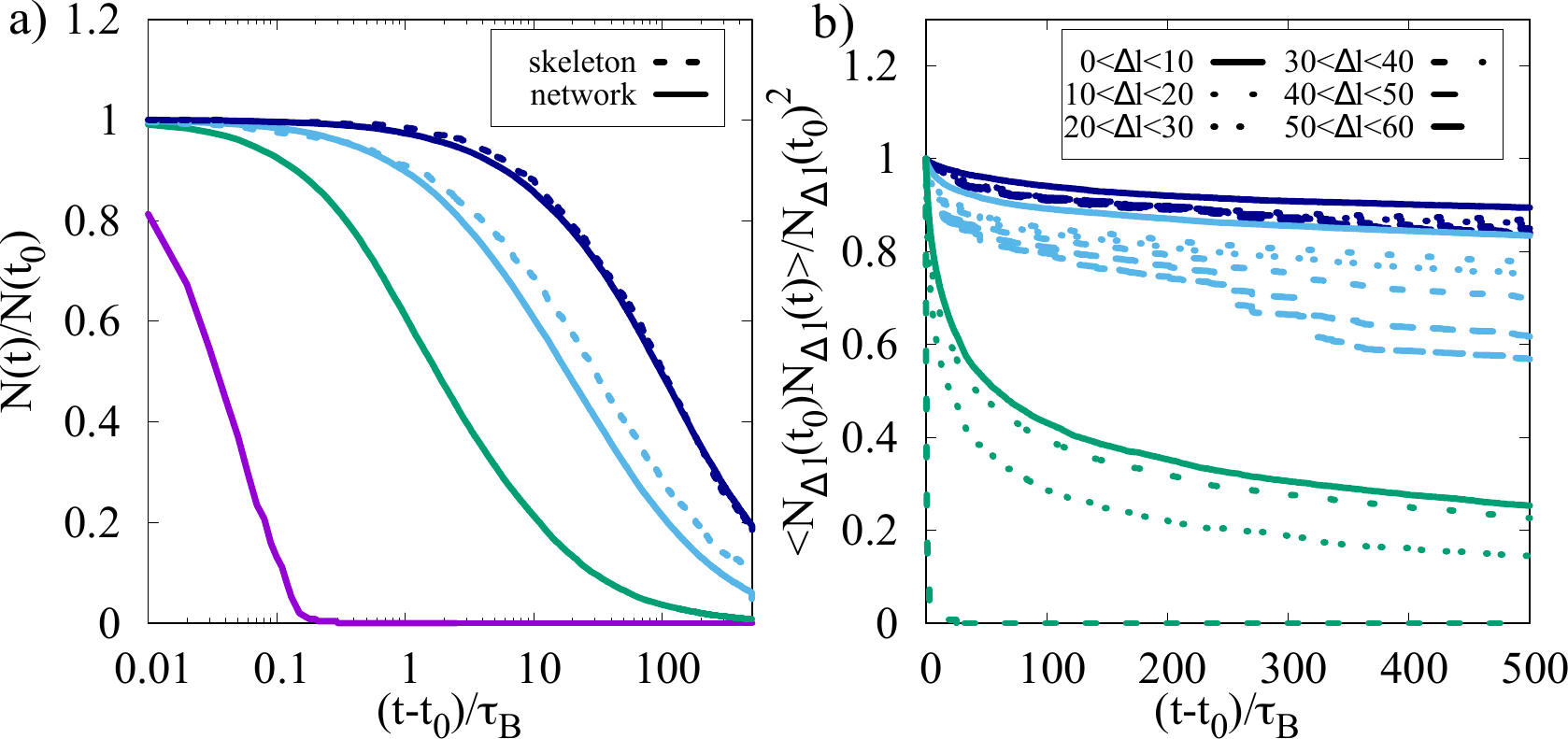}
\caption{\label{fig3.3.2-1} (a) Fractions of particles that stayed in
contact until time $t$ given that they were in contact at time $t_0$. The
colors indicate the strength of attraction as given in the legend in
Fig. \ref{fig3.2-1}. For the gel networks the result is also shown for
the cases where only particles in the reduced networks (skeleton) are
considered. (b) Correlation functions similar to those in Fig.
\ref{fig3.3.1-1}(a) where all real space distances $\Delta r$ are now
replaced by network distances $\Delta l$, i.e., the number of colloids
in the minimal path along particles in contact that connects the two
considered colloids. The same colors as in Fig. \ref{fig3.3.1-1}(a) are
used.}
\end{figure}

In Fig. \ref{fig3.3.2-1}(a) we show how many bonds between particles
survive up to a time $t-t_0$. As expected, bonds are more stable in case
of stronger attractions. Furthermore, in case of gel networks the bonds
between particles of the reduced network (dashed lines) are more stable
than bonds that occur anywhere (solid lines). Note that the differences
between particles of the reduced network and all particles of the
structure are more pronounced for intermediate attractions $\epsilon=9$
(light-blue) while they are small for the case of strong attractions
$\epsilon=15$ (dark-blue).

\subsubsection{Network correlations}
\label{sec3.3.2}

While the analyzes of the previous subsection only acknowledges the
rearrangements in space, we also want to explore changes of the network
topology. Therefore, we consider the network distance $\Delta l$ between
two particles that is defined as the number of particles in the minimal
chain of particles in contact that connects the initial particle with
the target particle. We calculate these distances for all pairs of
particles by using the Floyd-Warshall algorithm\cite{floyd,warshall}.

In the following we consider similar correlation functions as in the
previous subsection but the real space distances $\Delta r$ are now
replaced by network distances $\Delta l$. The resulting correlation
functions are shown in Fig. \ref{fig3.3.2-1}(b). In the case of the
cluster fluid with weak attractions $\epsilon=6.5$ the correlation
function decays very fast, while for the gel networks with $\epsilon=9$
and $\epsilon=15$ the decay is slower. Note that while the spatial
correlations in Fig. \ref{fig3.3.1-1}(a) were very similar for the two
considered gel networks, the correlations with respect to the network
topology in Fig. \ref{fig3.3.2-1}(b) differ significantly. The network
topology of gel networks with $\epsilon=15$ is more stable than the one
with $\epsilon=9$. That means once a suitable network configuration is
found the reformation of the network structure in a network with large
attraction strength is slower than for intermediate attractions. Note
that as we will show in the following even small rearrangements in space
can have a major impact on the network topology.

\begin{figure}
\includegraphics[width=\columnwidth]{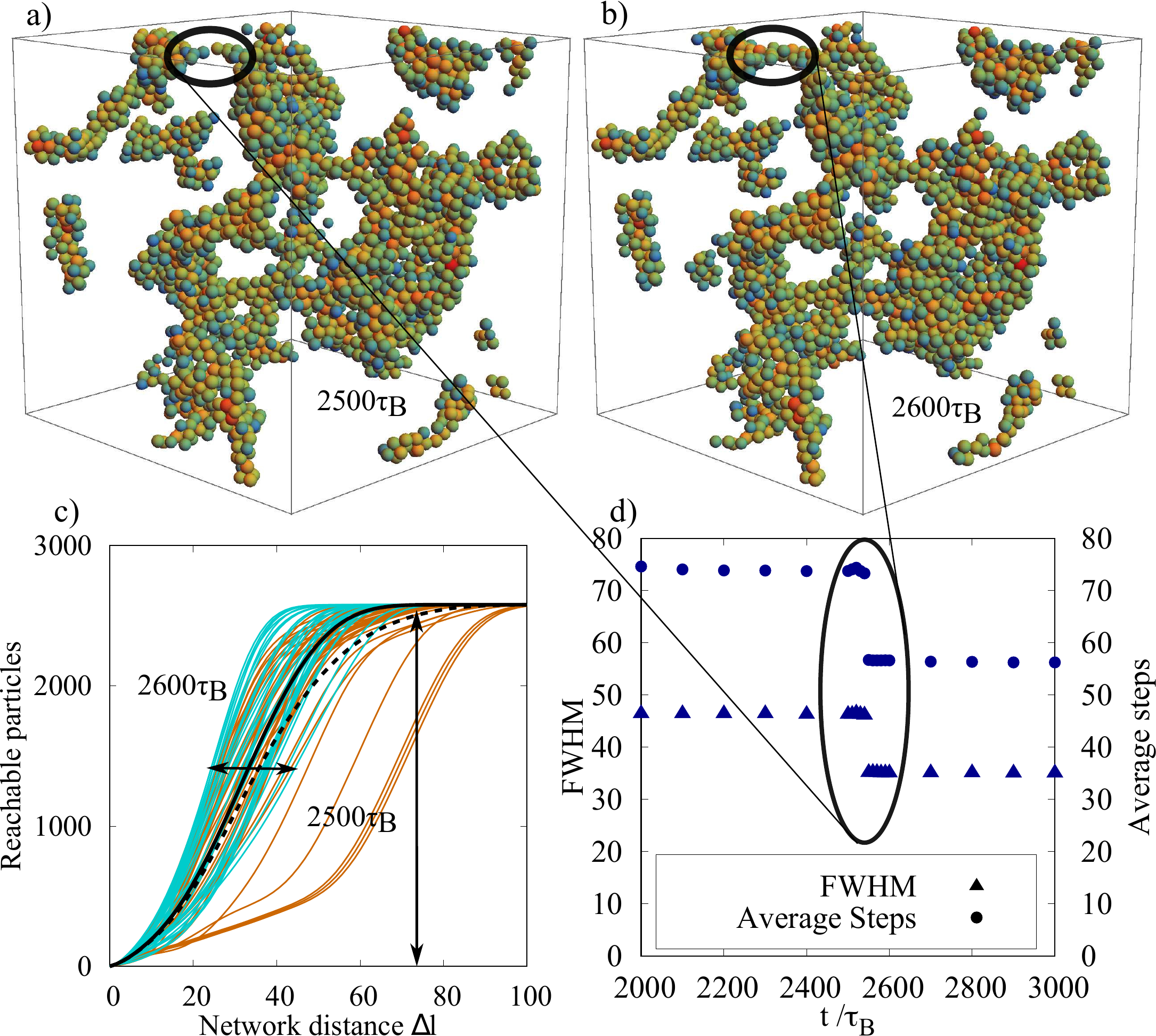}
\caption{\label{fig3.3.2-3} (a,b) Snapshots of a system with
$\epsilon=15$, $\kappa=15$, and $\eta=0.05$ after 2500$\tau_B$ and
2600$\tau_B$. A new connection forms in the top of the pictures. (c)
Number of particles that can be reached within $\Delta l$ steps along
particles in contact from a randomly chosen starting particle. For
visibility only curves for 30 randomly chosen starting particles are
depicted. If all possible 2578 curves were shown they would lay dense.
The brown curves are determined for the configuration after 2500$\tau_B$
depicted in (a) and the blue curves of the configuration after
2600$\tau_B$ shown in (b). The black lines show average curves (dashed
for 2500$\tau_B$, solid for 2600$\tau_B$). Note that some lines for
2500$\tau_B$ are hidden behind the curves for 2600$\tau_B$. (d) Triangles:
Full width at half maximum (FWHM) of the neighbor string plots as shown
in (c) where now all possible curves are considered in the evaluation.
Circles: Average number of steps needed to reach every particle in the
network. Both quantities are plotted as a function of time. The
connection event that is visible in the snapshots (a,b) leads to a jump
of the shown quantities.}
\end{figure}

In Fig. \ref{fig3.3.2-3} we depict how the appearance of a new
connection influences the network topology. It is clearly visible that
between the shown snapshots for $2500\tau_B$ and for $2600\tau_B$ a new
connection is formed (encircled in black). If we calculate for each
particle how many other particles can be reached in $k$ steps for a
randomly chosen starting particles we find curves as plotted in Fig.
\ref{fig3.3.2-3}(c). The range where these curves can be found indicates
that the new connection makes the network more dense in the sense that a
higher number of particles can be reached in less steps. To further
quantify this result we calculate the average number of steps needed to
reach the whole network and the full width at half maximum (FWHM) of the
curves as in Fig. \ref{fig3.3.2-3}(c) as a function of time. Fig.
\ref{fig3.3.2-3}(d) shows that between 2500$\tau_B$ and 2600$\tau_B$
these quantities jump to a lower value as expected because particles can
be reached faster and the density of the network increases.

These rearrangements of connections - either new formation or
destruction - also explain the small jumps that are visible in Fig.
\ref{fig3.3.2-1}(b) for $\epsilon=9$. Here either new connections are
formed or old connections are lost, such that big groups of particles do
not anymore belong to the same network distance range because the
topology of the network has changed dramatically.

\subsubsection{Relation between network distance and spatial distance}
\label{sec3.3.3}

\begin{figure}
\includegraphics[width=\columnwidth]{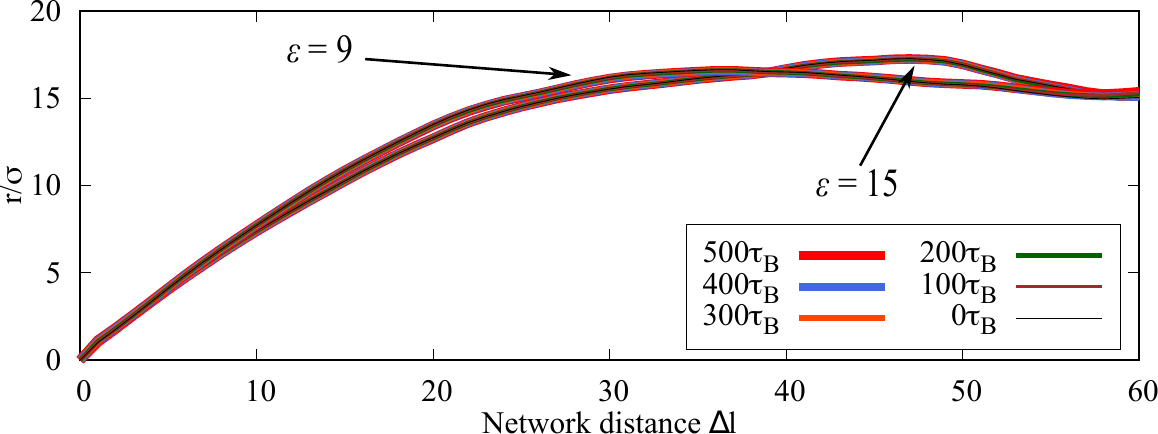}
\caption{\label{fig3.3.3-1} Spatial distance $\Delta r$ as a function of
network distance $\Delta l$ for gel networks with $\epsilon=9$ and
$\epsilon=15$. Otherwise, as always $\kappa =15$ and $\eta=0.05$. Ageing
times after an initial thermalization of $3000\tau_B$ are depicted by
the color. Note that for a given $epsilon$ the curves at different
ageing times collapse almost perfectly. Results are averaged over 4
independent runs.}
\end{figure}

We analyze the relationship between network distance and spatial
distance in gel networks. After an equilibration time of $3000\tau_B$ we
start from the meta-stable state which is reached and calculate average
spatial distance as a function of network distance. Fig.
\ref{fig3.3.3-1} shows the results for different ageing times after the
initial relaxation. For a given $\epsilon$ the curves at different ageing
times collapse almost perfectly indicating that the relation between
network topology and location in space is hardly affected by the ageing
process. Interestingly, there are differences for different attraction
strengths: The spatial distance shows a maximal value which depends on
the attraction strength. This indicates that for a large attraction of
$\epsilon=15$ longer and thinner strands form such that more steps are
needed to reach the same spatial distance. This finding is in agreement
with the larger mesh size observed in Fig. \ref{fig3.2-1}.
\section{Conclusion}
\label{sec4}
Due to the competition of depletion attractions and longer-ranged
repulsions between the colloidal particles in colloid-polymer mixtures,
complex, heterogeneous structures can be observed for attractions that
are stronger than at the binodal line of the fluid-liquid phase
separation in equilibrium. Our
Brownian dynamics simulations reveal significant differences between
unpercolated cluster fluids, gel networks, and clumpy gels. While the
dynamics of cluster fluids can be well described by free diffusion-like
motion of the clusters, the dynamics of gel networks and bulky gels is
much slower and does not reach a diffusive regime at the timescales that
are accessible to simulations.

Our structural analyses demonstrate that there are large voids in gel
networks. A typical length can be extracted resembling a mesh length.
Furthermore, we introduce reduced networks where all particles are left
out except for those that resemble the important connections between
particles. The reduced networks can be employed to identify percolating
strands and as an additional method to characterize the difference
between gel networks and clumpy gels.

The observed gel-like structures are meta-stable and the perfect
equilibrium states are actually still unknown (see also discussion in
\cite{archer}). We observe and characterize the ageing dynamics in all
gel-like states. We want to stress that different ageing processes can
be observed on different length scales, e.g., neighbor particles can
disconnect and connect again especially in case of smaller attractions.
On a longer length scale, namely typically the mesh size, the strands of
a gel network can slowly rearrange.

Concerning different gel networks, we find that correlations depending
on the network topology decay much faster in case of small attraction
then in case of strong attraction between the colloids. The reason for
this difference are strands that can rupture for small attractions but
hardly for strong ones. We observe that such rupture events do not have
an significant impact on the structure in space or the decay times of
spatial correlations. However, we expect that the rupture of strands
might influence the mechanic stability or the shear slabs that occur
under continuous shear \cite{kohl2} which will be the topic of future works.

Furthermore, in future we also want to explore the slowdown of the
different types of dynamics in a gel network in more detail. In a clumpy
gel at large densities the slowdown sometimes is attributed to a
glass-like effective ergodicity breaking \cite{toledano,zhang}, we
expect that that the different types of dynamics that occur in a gel
network become arrested at different transitions. As we have related the
ergodicity breaking transition of thermal jamming to a directed
percolation in time \cite{milz,maiti}, we want to find out whether the
step-wise dynamic arrest in gel networks also can be related to
percolation transitions in time and how these are related to the
percolation in space.


%
%

%


\section*{Data Availability Statement}
The data that support the findings of this study are available from the corresponding author upon reasonable request.
\begin{acknowledgments}
We want to thank S. Egelhaaf for useful discussions.
\end{acknowledgments}


\end{document}